# A Model of Optimal Network Structure for Decentralized Nearest Neighbor Search

Alexander Ponomarenko, Irina Utkina, Mikhail Batsyn

National Research University Higher School of Economics
aponomarenko@hse.ru, iutkina@hse.ru, mbatsyn@hse.ru

**Abstract**
One of the approaches for the nearest neighbor search problem is to build a network which nodes correspond to the given set of indexed objects. In this case the search of the closest object can be thought as a search of a node in a network. A procedure in a network is called decentralized if it uses only local information about visited nodes and its neighbors. Networks, which structure allows efficient performing the nearest neighbor search by a decentralized search procedure started from any node, are of particular interest especially for pure distributed systems. Several algorithms that construct such networks have been proposed in literature. However, the following questions arise: "Are there network models in which decentralized search can be performed faster?"; "What are the optimal networks for the decentralized search?"; "What are their properties?". In this paper we partially give answers to these questions. We propose a mathematical programming model for the problem of determining an optimal network structure for decentralized nearest neighbor search. We have found an exact solution for a regular lattice of size 4x4 and heuristic solutions for sizes from 5x5 to 7x7. As a distance function we use $L_1$, $L_2$ and $L_\infty$ metrics. We hope that our results and the proposed model will initiate study of optimal network structures for decentralized nearest neighbor search.

## 1 Introduction

The nearest neighbor search appears in many fields of computer science. A problem of building data structure for the nearest neighbor search is formulated as follows. Let $D$ be a domain and $d: D \times D \to R_{[0;+\infty)}$ be a distance function. One needs to preprocess a finite set $X \subseteq D$ so that the search of the closest object for any given query $q \in D$ in the set $X$ will be as fast as possible. A huge number of methods have been proposed. Of particular interest is the case when the search of nearest neighbor should run in a distributed environment without any central coordination point. For this case a natural approach for organizing nearest neighbor search is to build a network, which nodes correspond to the given set $X$. In this case the search of the closest object can be thought as a search of a node in a network. Moreover a distributed environment, especially for p2p case, requires that all procedures that are involved in the search or indexing processes should be decentralized. This means that all procedures have only local information about visited nodes and its neighbors and don't have access to the information about the whole structure of the network.

As a rule such an approach implies searching via *greedy walk* algorithm [1,3,7,8] or its modification [6,9]. So, many p2p systems including DHT protocols [2,4,5] use the same search algorithm, but employ different distance functions and have different network structures.

In the present paper we address the problem of optimal network structure for NNS. We emphasize that for any fixed input set there exists an optimal network structure with respect to the chosen search algorithm. To study the properties of such networks, we present a mathematical Boolean non-linear programming model of optimal network structure. The objective is to minimize the expected number distance computations made by the greedy walk algorithm to find the nearest neighbor for an arbitrary query starting from an arbitrary node.

As a first step we solve this problem for the case when the input set $X$ corresponds to the set of nodes of a two-dimensional regular lattice. We have found an exact solution for size 4x4 and heuristic solutions for sizes from 5x5 to 7x7. As a distance function we use $L_1$, $L_2$ and $L_\infty$ metrics.

## 2  Mathematical formulation

We consider a network as a graph $G(V,E)$ with vertex set $V = X = \{1,...,n\}$ and edge set $E \subset V \times V$. Let $d(i,q)$ be a distance function between vertex $i$ and query $q$. The neighborhood of vertex $i$ is defined as $N(i) = \{j \in V : (i,j) \in E\}$. We denote the probability function for a query as $f_q$ for a discrete domain and as $f(q)$ - the probability density function for a query in continuous domain.

### 2.1  Decentralized Search Algorithm - Greedy Walk

The goal of the search algorithm is to find the vertex (target vertex) in the graph $G$ which is the closest to the query, going from one vertex to another through the set of edges $E$ of $G$. The search is based on the information related to the vertices. During the search process the algorithm can calculate the distance between the query and the vertices which it knows. Below is the pseudo code of the greedy walk algorithm.

**GreedyWalk(** $s \in V$, $q \in V$ **)//** $s$ -starting vertex, $q$ - query
1    $c \leftarrow \underset{y \in N(s)}{\mathrm{argmin}}(d(y,q))$
2    **if** $d(c,q) < d(s,q)$ **then**

```
3       return GreedyWalk(c, q)
4   else
5       return s
```

Starting from vertex $s$ the algorithm calculates the value of the distance function $d(y,q)$ between query $q$ and every neighbor $y$ of $s$. After that the algorithm is recursively called for vertex $c$ closest to the $q$. The algorithm stops at the vertex which neighborhood contains no vertices closer to the query than itself. The greedy walk algorithm can be also considered as a process of routing a search message in a network. At each step the node (vertex) which has received a message (message holder) passes it to the neighbor closest to the query according to the function $d$.

### 2.2  Mathematical programming model

By no means all graphs have proper structure for searching via greedy walk. In our model we require from the structure of graph $G$ that search of any vertex by the greedy walk will reach the target vertex starting from an arbitrary vertex. In general this requires that the graph need to have the Delone graph as a subgraph. Similar to the Kleinberg model [1] in this paper we consider a particular case when vertices are nodes of a regular lattice with integer coordinates. In this case the Delone graph is just the set of the edges of the regular lattice.

The complexity of the search algorithm is measured as the number of different vertices for which the distance to the query has been calculated. We take this number as an objective function. Equations (1-9) define Boolean non-linear programming formulation for optimal graph structure.

Decision variables

$$x_{ij} = \begin{cases} 1, & \text{if edge } (i,j) \text{ belongs to the solution} \\ 0, & \text{otherwise} \end{cases} \quad (1)$$

$$y_{ij}^k = \begin{cases} 1, & \text{if vertex } k \text{ belongs to the greedy walk from } i \text{ to } j \\ 0, & \text{otherwise} \end{cases} \quad (2)$$

Objective function

$$\min \frac{1}{n} \sum_{i=1}^{n} \sum_{q \in D} O(i, j_q) f_q \text{ (discrete domain)} \quad (3a)$$

$$\min \frac{1}{n} \sum_{i=1}^{n} \int_{D} O(i, j_q) f(q) dq, \text{ (continuous domain)}, \quad (3b)$$

$$\text{where } j_q = \arg\min_{j=\overline{1,n}} d(j,q) \tag{4}$$

$$O(i,j_q) = \left|\left\{l \in V : \exists k \; x_{lk} = 1 \text{ and } y^k_{ij_q} = 1\right\}\right| \tag{5}$$

Constraints

$$x_{ii} = 0 \quad \forall i \in V \tag{6}$$

$$y^i_{ij} = y^j_{ij_q} = 1 \quad \forall i, j_q \in V \tag{7}$$

$$\sum_{k=1}^{n} x_{lk} y^k_{ij_q} \geq y^l_{ij_q} \quad \forall i, j_q, l \in V \tag{8}$$

$$l^* = \arg\min_{l \in V : x_{kl}=1}(d(l,q)) \implies y^{l^*}_{ij_q} \geq y^k_{ij_q} \quad \forall q \in D \; \forall i,k \in V \tag{9}$$

Decision variables $x_{ij}$ (1) determine the adjacency matrix of the optimal graph, which we want to find. Indicator variables $y^k_{ij}$ (2) are used to calculate the number of the operations $O(i,j_q)$ performed during the search process from vertex $i$ to vertex $j_q$, which is the closest vertex (target vertex) to the query $q$ (4). In our case it is the number of different vertices for which the distance to the query has been calculated. This is equal to the cardinality of the union set of neighborhoods of vertices $k$ for which $y^k_{ij} = 1$ (5).

Since we want to find the optimal graph in general case (for any starting vertex and any query) our objective is to minimize the average number of operations required for the search algorithm to reach a target vertex (3a, 3b, 4, 5). Constraint (6) guarantees that there are no loops in the graph and constraint (7) requires GreedyWalk($i, j$) to start from vertex $i$ and stop at vertex $j$. Constraint (8) links variables $x_{ij}$ and $y^k_{ij}$ and requires that the search algorithm (the greedy walk) will go through one of vertex $l$ neighbors if it goes through this vertex $l$. Constraint (9) describes the greedy strategy of the greedy walk algorithm: if vertex $k$ belongs to the greedy walk from vertex $i$ to vertex $j_q$ ($y^k_{ij_q} = 1$) then its neighbor $l^*$, closest to the query $q$ among all its neighbors $l$, should also belong to this greedy walk ($y^{l^*}_{ij_q} = 1$).

The presented model is applicable for an arbitrary metric space. In the next section we present the results for a particular case when vertices are the nodes of a two-dimensional regular lattice and the distance functions are $L_1$, $L_2$, or $L_\infty$.

## 3  Computational Experiments and Results

In this work we suppose that the input set corresponds to the nodes of a two-dimensional regular lattice and we have a domain such that all nodes have the same probability to be the nearest neighbor for a query. In this case the nearest neighbor search can be thought as a node discovery procedure, which means that we need to find the given node in the network.

Obviously, we can find the optimal graph structure if we check all possible configurations of the set of edges. However the number of all possible configurations grows as $2^{n(n-1)/2}$.

To find an exact solution we have implemented a branch and bound algorithm. The exact solutions found by algorithm for regular lattice 4x4 are presented at Fig. 1. The solutions found by our heuristic are presented at Fig. 2-4.

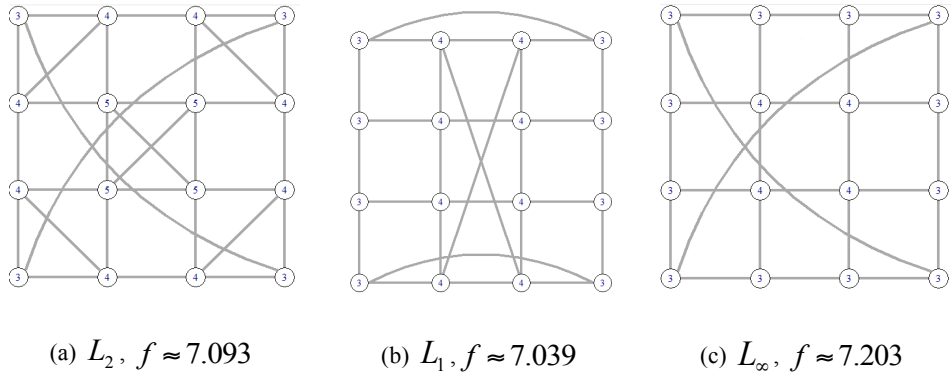

(a) $L_2$, $f \approx 7.093$      (b) $L_1$, $f \approx 7.039$      (c) $L_\infty$, $f \approx 7.203$

Fig. 1. Exact solutions found by our branch and bound algorithm for regular lattice 4x4

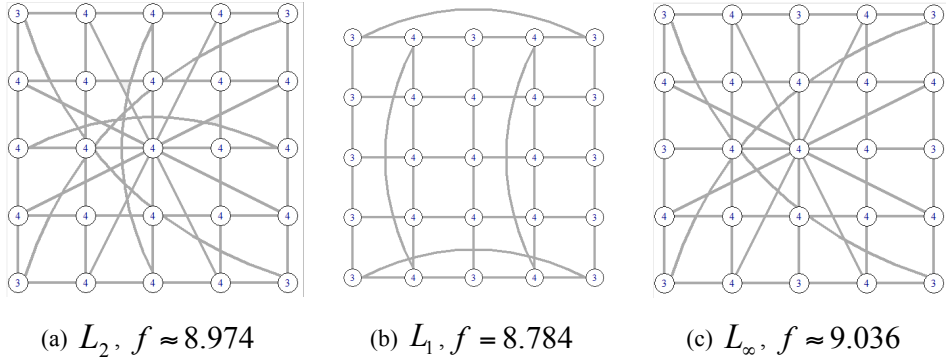

(a) $L_2$, $f \approx 8.974$    (b) $L_1$, $f = 8.784$    (c) $L_\infty$, $f \approx 9.036$

Fig. 2. Solutions found by our heuristic for regular lattice 5x5

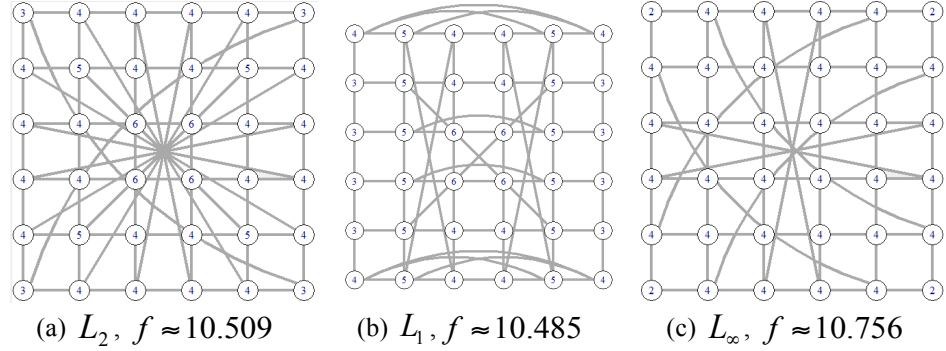

(a) $L_2$, $f \approx 10.509$    (b) $L_1$, $f \approx 10.485$    (c) $L_\infty$, $f \approx 10.756$

Fig 3. Solutions founded by heuristic for a regular lattice 6x6

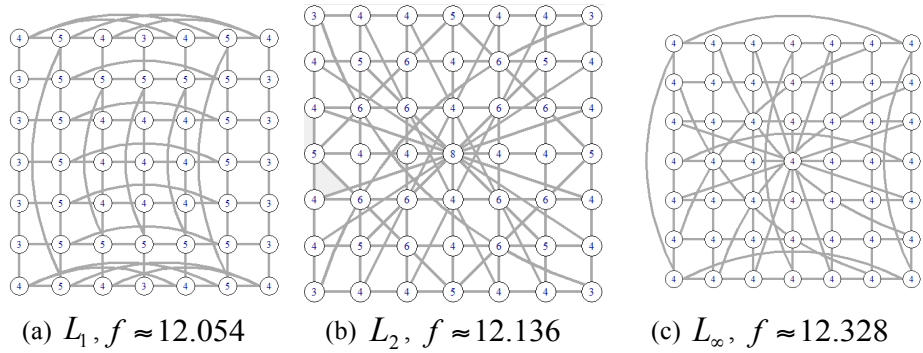

(a) $L_1$, $f \approx 12.054$    (b) $L_2$, $f \approx 12.136$    (c) $L_\infty$, $f \approx 12.328$

Fig. 4. Solutions found by our heuristic for regular lattice 7x7

## 4 Conclusion and Future Work

We have proposed a Boolean non-linear programming model to determine an optimal graph structure, which minimizes the complexity of the nearest neighbor search by the greedy walk algorithm. We have found an exact solution for a regular lattice of size 4x4 and presented the results found by our heuristic for sizes from 5x5 to 7x7 with the three most popular distances: $L_1$, $L_2$ and $L_\infty$.

However, we realize that the most important characteristic which should be studied is the asymptotical behavior of the objective function. Therefore our future work will be focused on improving the efficiency of our exact and heuristic algorithms. We also have plans to develop models describing optimal network structures for approximate nearest neighbor search. We hope that this work will draw attention to the study of graph structures optimal for decentralized nearest neighbor search.

## Acknowledgments

This research is conducted in LATNA Laboratory, National Research University Higher School of Economics and supported by RSF grant 14-41-00039.